\documentclass[useAMS,usenatbib]{mn2e}
\usepackage{graphicx}
\usepackage{aas_macros}

\title[DNOs and QPOs in cataclysmic variables -- VI]{Dwarf nova oscillations and quasi-periodic oscillations in cataclysmic variables -- VI. Spin rates, propellering, and coherence}
\author[B. Warner and M.L. Pretorius]{Brian Warner$^{1,2}$\thanks{E-mail: warner@physci.uct.ac.za (BW); mlp@astro.soton.ac.uk (MLP)} and Magaretha L. Pretorius,$^{2}$\footnotemark[1]\\
$^{1}$Department of Astronomy, University of Cape Town, Rondebosch 7700, Cape Town, South Africa\\
$^{2}$School of Physics and Astronomy, University of Southampton, Highfield, Southampton SO17 1BJ, United Kingdom\\}

\begin{document}

\date{Accepted . Received ; in original form }

\pagerange{\pageref{firstpage}--\pageref{lastpage}} \pubyear{}

\maketitle

\label{firstpage}

\begin{abstract}
We examine published observations of dwarf nova oscillations (DNOs) on the rise and decline of outbursts and show that their rates of change are in reasonable agreement with those predicted from the magnetic accretion model. We find evidence for propellering in the late stages of outburst of several dwarf novae, as shown by reductions in \emph{EUVE} fluxes and from rapid increases of the DNO periods. Reanalysis of DNOs observed in TY PsA, which had particularly large amplitudes, shows that the apparent loss of coherence during late decline is better described as a regular switching between two nearby periods. It is partly this and the rapid deceleration in some systems that make the DNOs harder to detect.
   
We suggest that the 28.95~s periodicity in WZ Sge, which has long been a puzzle, is caused by heated regions in the disc, just beyond the corotation radius, which are a consequence of magnetic coupling between the primary and gas in the accretion disc. This leads to a possible new interpretation of the `longer period DNOs' (lpDNOs) commonly observed in dwarf novae and nova-like variables.
\end{abstract}


\section{Introduction}
Dwarf nova oscillations (DNOs) were first recognized in outbursts of dwarf novae and in nova-like variables; that is, during stages of high mass transfer rates ($\dot{M}$) from companion to the white dwarf primary of cataclysmic variables (Warner \& Robinson 1972). Their short time scales ($\sim 5$--50~s) indicate originations close to the surface of the primary, but their large variations in period distinguish them from the short period highly stable magnetic accretors such as DQ Her and AE Aqr (for a general review see Warner 1995a). Paczynski (1978) suggested that these variations result from the torque acting on an accreted equatorial belt on the primary during phases of changing $\dot{M}$. This model has been extended in Warner (1995b) and the earlier papers in this series (Woudt \& Warner 2002, Warner \& Woudt 2002, 2006; Warner, Woudt \& Pretorius 2003; Pretorius, Warner \& Woudt 2006); a review is given in Warner (2004).

In this contribution we reinvestigate some early observations, in the light of later developments, looking specifically for additional evidence relevant to the Low Inertia Magnetic Accretor (LIMA) model (Warner \& Woudt 2002; hereafter WW2002). In Section 2 we compare predicted and observed rates of change of spin on the rise and fall of dwarf nova outbursts; in Section 3 we look at possible incidences of a propellering phase late in dwarf nova outbursts; in Section 4 we consider the question of `coherence' of DNOs; in Section 5 we suggest a model for the multiple DNOs in WZ Sge, their behaviour during outburst, and their possible relevance to the distinct group of longer period DNOs (lpDNOs). Section 6 contains a short general conclusion.

\section{Spin-up and spin-down}
Our concept of the progress of DNOs through a dwarf nova outburst is that, at the beginning of outburst, material moving inwards and reaching the inner parts of the accretion disc is either magnetically channeled into the equatorial regions of the primary or reaches the equator without channeling, depending on the magnetic field in the inner regions. Even if the field of the primary is quite small, differential rotation of the ensuing equatorial belt can use this seed field and may amplify it to the point where channeling can occur (WW2002). The accretion zones generated by channeling are directly or indirectly responsible for the DNOs. Below some critical field of the primary (estimated in WW2002 to be $\sim 10^4\,\mathrm{G}$) there will be no channeling and no DNOs during outburst. This is why otherwise very similar systems can exhibit completely different behaviours of their DNOs (e.g. VW Hyi and EK TrA; Warner \& Woudt 2006).

Moreover, for primary fields less than $\sim 10^5\,\mathrm{G}$ the outer regions are not coupled to the interior of the primary and so the accreted gas is free to rotate as an equatorial belt as it builds up in mass during the outburst. It is, however, magnetically coupled to the disc in the same way as in intermediate polars (IPs), and therefore as the inner edge of the disc moves inwards through the increased thrust of the mass inflow, accessing shorter periods of rotation, the belt is spun up, and vice versa after maximum of outburst. Through the maximum phases of outburst the accretion is near equilibrium disc accretion, as is evidenced by the higher stability (e.g. observed low $\dot{P}$) of the DNOs during those times.
   
We here extend the quantitative aspects of this model, as developed in WW2002: we note that from the standard theory of accretion from a disc into a magnetosphere (e.g. Ghosh \& Lamb 1979; Chapter 7 of Warner 1995a) the angular acceleration generated by the combined material and magnetic torques acting on an equatorial belt with mass $M(b)$, given in equation (5) of WW2002, can be written in terms of the period of rotation $P$ of the belt as
\begin{equation}
\dot{P}=2.67 \times 10^{-8}n(\omega_s)\omega_s^{1/3}\dot{M}_{16}M_{22}^{-1}(b)M_1^{4/3}(1)P^{7/3}
\end{equation}
where $n(\omega_s)$ is the torque function, $\omega_s$ is the ``fastness parameter'', given by $r_0 = \omega_s^{2/3}r_{co}$, where $r_0$ is the radius of the inner edge of the disc and $r_{co}$ is the corotation radius, found from $r_{co}^3=GM(1)/\Omega^2(b)$, $M(b)$ is the mass of the belt, $\Omega(b)$ is its angular velocity, $\dot{M}$ is the rate of mass transfer onto the primary, and $M(1)$ is the mass of the primary. Subscripts denote cgs values in powers of ten, with the exception that $M_1$ is used for mass in solar units.

An outburst can be partitioned into two sets of conditions: on the rise and on the fall the accretion process is not close to equilibrium, i.e. the rotation periods of the belt and the inner edge of the accretion disc are significantly different, but around maximum the accretion is close to equilibrium, with the result that $n(\omega_s) \sim 0$, $\omega_s \sim 0.975$ and $\dot{P} \sim 0$.  On the rise and fall, however, we have $n(\omega_s) \sim 0.7$ (Wang 1987) and $0.25 \le \omega_s \le 1.0$, so the factor $\omega_s^{1/3}$ in equation (1) is bounded by $0.63 \le \omega_s^{1/3} \le 1.0$; the upper bound comes from the requirement that gas threading onto field lines must lie within the corotation radius, and the lower bound arises from (a) the requirement that $r_0 \ge R(1)$ and (b) the observed low values of $P$ for the DNOs that imply that the inner edge of the disc is never very far from the surface of the primary.

Next we note that $\dot{M}$ during the outburst phases of interest here probably does not vary greatly. The models of accretion during outburst produced by Cannizzo (1993) show that, from the mid-point of optical rise to the midpoint of optical decline, $\dot{M}$ onto the primary varies only about a factor of two about the average. These correspond roughly to the phases over which most DNOs are observed. Therefore the mass of the equatorial belt (which we assume to be very small at the outset) during the interesting parts of the outburst can be approximated by $\dot{M}(b) \propto T$, where $T$ is the time since the start of outburst.

For the two non-equilibrium phases we therefore have (with appropriate choices of sign) 
\begin{equation}
\dot{P} \simeq \pm 1.6 \times 10^{-4}\left[ M_1(1)/0.6 \right]^{4/3}\left(P/25\,\,\mathrm{s}\right)^{7/3}T^{-1}
\end{equation}
where $T$ is measured in days.

We therefore expect $\dot{P} \sim 1 \times 10^{-4}$ on the rise and $\dot{P} \sim 3 \times 10^{-5}$ on the fall of a normal dwarf nova outburst.  In Tables 1 and 2 we tabulate values of $\dot{P}$ deduced from published observations.  There are relatively few usable observations: for the rise this is because very few dwarf novae have been caught early on the rise, and for the decline it is because the DNOs often become too `incoherent' to detect with Fourier transforms late in outburst. As an example, VW Hyi is one of the most intensively studied systems (see list of observations in Woudt and Warner 2002) but has never been caught showing DNOs on the rise, and there is only one detection of a DNO on the decline before the rapid deceleration phase begins, as described below. And we need to avoid the late rise or early fall where the low $\dot{P}$ characteristic of maximum still has influence. However, it is immediately obvious from Tables 1 and 2 that $|\dot{P}|$ is roughly a factor of five or more larger on the rise than in mid-decline. In the LIMA model this is expected because of  the smaller inertia of the equatorial belt in the rising phase, compared with that of the almost fully accreted mass late in decline. Furthermore, after allowance for the range of parameters (e.g. $M_1(1) \sim 1.2$ and $P \sim 10\,\mathrm{s}$ in SS Cyg, which have opposite influences in equation (2) -- and this is a general rule, the higher mass primaries will generate shorter period DNOs) there is moderately good agreement with the predictions of equation (2). This is new evidence for the viability of the LIMA model.
    
The notably larger values of $\dot{P}$ in Table~2, observed for OY Car, SS Cyg and TY PsA \emph{late} in decline, are discussed in the next Section, along with a similar effect seen in VW Hyi.

\begin{table*}
 \centering
 \begin{minipage}{110mm}
  \caption{DNOs on early rise.  The fifth column lists the range in time over which the period evolution given in the fourth column occurred.}
  \begin{tabular}{lllllll}
  \hline
Star    & Outburst & $T$ (d) & $P_{DNO}$ (s) & Range (d) & $\dot{P}/10^{-5}$ & References \\
  \hline
SS Cyg  & Jun 1984 & 1.9            & 10.1 -- 9.2          & 0.19               & 8                 & 1 \\
SS Cyg  & Mar 1997 & 1              & 7.81 -- 6.59         & 0.20               & 7.1               & 2 \\
AH Her  & Jun 1976 & 2              & 31.3 -- 30.7         & 0.1                & 7                 & 3 \\
AH Her  & Jun 1978 & 2 -- 3         & 27.8 -- 25.0         & 1.0                & 3.2               & 3 \\
SY Cnc  & Jan 1975 & 2 -- 3         & 29.50 -- 26.71       & 1.0                & 3.2               & 3 \\
  \hline
  \end{tabular}
References: 1. Jones \& Watson 1992; 2. Mauche \& Robinson 2001; 3. Patterson 1981.\hfill
\end{minipage}
\end{table*}

\begin{table*}
 \centering
 \begin{minipage}{120mm}
  \caption{DNOs during mid- and late-decline.  The fifth column lists the range in time over which the period evolution given in the fourth column occurred.}
  \begin{tabular}{lllllll}
  \hline
Star    & Outburst & $T$ (d) & $P_{DNO}$ (s) & Range (d) & $\dot{P}/10^{-5}$ & References \\
  \hline
AH Her  & Jun 1978     & 7 -- 10    & 24.25 -- 25.75       & 3.0                & 0.58              & 1 \\
SY Cnc  & Jan 1975     & 8          & 27.80 -- 29.08       & 1.0                & 1.5               & 1 \\
        &              & 9          & 29.08 -- 31.20       & 1.0                & 2.5               & 1 \\
OY Car &Nov 1980$^\ast$& 12:        & 19.44 -- 20.48       & 1.0                & 1.2               & 2 \\
       &               & {\it 13:}  &{\it 20.48 -- 27.99}  &{\it 1.0}           &{\it 8.7}          & 2 \\  
TY PsA &Jun 1984$^\ast$& 10 -- 13   & 25.2 -- 30.2         & 3.0                & 1.9               & 3 \\
       &               &{\it 13 -- 14} &{\it 30.2 -- 38.0} &{\it 1.0}           &{\it 9.0}          & 3 \\
       &               &{\it 14 -- 15} &{\it 38.0 -- 47.0} &{\it 1.0}           &{\it 10.4}         & 3 \\
       & Sep 1984      &{\it 1 -- 2}   &{\it 26.5 -- 36:}  &{\it 1.0}           &{\it 11:}          & 3 \\
V1159 Ori&Dec 1993$^\ast$& 10 -- 12 & 31.61 -- 32.21       & 2.0                & 0.35              & 4 \\
       &               & 12 -- 15   & 32.21 -- 33.92       & 3.0                & 0.66              & 4 \\
SS Cyg & Sep 1984 & 4               &  7.8                 &                    & 0.46              & 5 \\
       &          & 6               & 10.4                 &                    & 0.66              & 5 \\
       &          &{\it 8 or 9}        &{\it 10.7}            &{\it $\sim0.5$}     &{\it 6--10}        & 5 \\
  \hline
  \end{tabular}
Notes: $^\ast$ indicates a superoutburst. Italicized entries are for late-decline.\\
References: 1. Patterson 1981; 2. Schoembs 1986; 3. Warner, O'Donoghue \& Wargau 1989, and Section 4 of this paper; 4. Patterson et al. 1995; 5. Jones \& Watson 1992.\hfill
\end{minipage}
\end{table*}

\section{Propellering}
As the system passes through the later declining phases of the outburst, $\dot{M}$ is rapidly reduced, acting to increase the radius $r_0$ of the magnetosphere, but at the same time any sheer-induced magnetic field in the belt is declining, which acts to reduce $r_0$. The rate of increase or decrease of $r_0$ therefore depends on details of these changes, which are currently unknown. One possibility is that $r_0$ increases much faster than $r_{co}$ (which is determined by the spin period $P(b)$), catches up with it and turns an $\omega_s \le 1$ accretor into an $\omega_s \ge 1$ non-accretor. That is, gas will be centrifuged outwards on the field lines, in a propeller action, as has been suggested for VW Hyi (WW2002).
   
In VW Hyi, unlike other dwarf novae, the late decline shows DNOs of large amplitude. These have a rate of increase of period nearly two orders of magnitude larger than predicted by equation (2), showing that some additional physics is required. For VW Hyi, WW2002 found that the observed rate $\dot{P} = 7.5 \times 10^{-4}$, which is deduced from an increase of $P$ from 23~s to 90~s in 24 hours (Warner \& Woudt 2006; hereafter WW2006), may be explained by the angular momentum extracted from the equatorial belt as it feeds into the expelled gas. A concomitant precipitate drop in \emph{EUVE} flux during just this phase (Mauche, Mattei \& Bateson 2001: hereafter MMB; WW2002) is direct evidence for the reduction of mass falling onto the primary.
  
In Table~2 the large increase of $\dot{P}$ in the last observed DNOs in OY Car is direct evidence for rapid deceleration of the belt -- but no $EUV$ flux has been observed at this stage. Also, a large temporary reduction in $EUV$ flux from SS Cyg is seen late in outburst (MMB), and it is during this phase of outburst that Jones \& Watson (1992) found that a large increase in $\dot{P}$, rather than a decrease in phase coherence, can explain the range of observed DNO phases. The one very late phase $EUV$ flux observation in U Gem (MMB; Long et al. 1996) also shows a precipitate drop, starting midway down the optical decline as in VW Hyi, but no DNOs have been detected at this time. The increase of $\dot{P}$ after $T = 13$~d in TY PsA late in the superoutburst of June 1984, although not greatly in excess of what is predicted by equation (2), is similar to what is seen in OY Car. These results for the late declines of VW Hyi, SS Cyg, OY Car, U Gem and TY PsA are indications that propellering is a common event in late stages of at least some dwarf novae outbursts. 

Although there is not yet a complete theory of accretion onto the primary during the late stages of outburst we can provide an illustrative example of why the expansion of the magnetosphere may overtake the expansion of the corotation radius, leading to propellering. From the definition of the corotation radius we have
\begin{equation}
d r_{co}/d t = 1.00 \times 10^8 P^{-1/3} \dot{P} M_1^{1/3}(1)\,\mathrm{cm\,s^{-1}}.
\end{equation}
The radius $r_\mu$ of the magnetosphere, which we take here to be effectively $r_0$, is, for a dipole field, (e.g. Warner 1995)
\begin{equation}
r_0 = 5.15 \times 10^{10} \mu_{34}^{4/7} M_1^{-1/7}(1)\dot{M}_{16}^{-2/7}\,\mathrm{cm},
\end{equation}
where $\mu$ is the magnetic moment of the primary (or, in this case, the equatorial belt), i.e. $\mu = B(b)R^3(1)$, where $B(b)$ is the field in the belt. If we use the theory of field enhancement given in equation (14) of WW2002, and adopt appropriate values for the parameters, we find
$$B_3(b) \simeq 1.18 B_r^{1/2}(1) M_{22}^{1/4}(b) P^{-1/2} M_1^{-1/4}(1)\,\mathrm{G},$$
which leads to 
\begin{equation}
r_0 = 1.71 \times 10^8 B_r^{2/7}(1) M_{22}^{1/7}(b) P^{-2/7} \\
M_1^{-6/7}(1) \dot{M}_{16}^{-2/7}\,\mathrm{cm},
\end{equation}
%
which can be differentiated to provide $dr_0/dt$. Before progressing further, though, we need an estimate of $\dot{M}$ and in particular the way it varies in the late decline stages of outburst. There is as yet no theory for this, but in the case of VW Hyi we can use the $EUV$ flux, before propellering starts, as a proxy for $\dot{M}$. From MMB (see fig.~1 of WW2002) we find $\dot{M} \propto \exp \left(-t/0.36\,\mathrm{d}\right)$.  Adopting $M_1(1)=0.6$, $\dot{M}_{16} \simeq 5$, $M_{22}(b)=5$, $ B_r(1)=10^4\,\mathrm{G}$ and $\dot{P}$ from eq. (2), then eq. (5) gives $dr_0/dt \simeq 2.2 \times 10^9\,\mathrm{cm\,d^{-1}}$ and eq. (3) gives $dr_{co}/dt \simeq 3 \times 10^8\,\mathrm{cm\,d^{-1}}$.  These rough estimates indicate that propellering can indeed be a consequence of the rapid drop in $\dot{M}$ near the end of outburst. The propellering will continue until the equatorial belt is decelerated to the point where equilibrium accretion from the inner edge of the disc is re-established.
   
As shown by Spruit \& Taam (1993; hereafter ST), unless $\omega_s$ is very large, very little gas is ejected from the system -– it is merely moved outward in the disc and will be accreted later. In the case of VW Hyi, accretion at a low level ($\dot{M}_{16} \sim 1$) is frustrated for about 24 h while it is being centrifugally expelled by magnetic coupling to a belt that was accumulated during several days at $\dot{M}_{16} \sim 50$; the angular momentum drain requires only that the rejected gas be moved out to $\sim 10 R(1)$.  The analysis by ST shows that diffusion in the disc brings some of the expelled gas back to the inner edge, with the result that the increasing pressure there reduces the radius of the inner edge and the condition $\omega_s < 1$ is eventually regained. In the case of dwarf novae this is not a cyclical process (as investigated by ST) because the inner disc is rapidly drained of mass after the propellering ceases.

\section{Coherence of DNOs}
From the earliest years of their study the DNOs have been recognized to have variations in period or phase that occur on short times scales, as well as through an outburst. Superimposed on the general spin-up and spin-down are sudden small changes of period. In general the DNOs are more stable at maximum of outburst, and deteriorate in stability as the outburst progresses -- this was first made quantitative by Hildebrand, Spillar \& Stiening (1981) from optical measurements and by Cordova et al. (1984) from X-ray observations. The disappearance of DNOs (if present at all) from Fourier transforms of light curves in the late stages of dwarf nova outbursts has been attributed to this effect.
  
In the absence of a physical model for DNOs it was a natural course to examine their behaviour by statistical analyses (see Chapter 8 of Warner 1995a for an overview) and comparison with particular models. These included noise-excited damped harmonic oscillator (Robinson \& Nather 1979) and random walk in phase of a sinusoidal oscillator (Horne \& Gomer 1980; Cordova et al. 1984). These were the best that could be achieved for the low amplitude oscillations, but examination of the few examples of large amplitude oscillations has repeatedly thrown doubt on the value of such an approach. Relevant statements are: (referring to phase measurements during an outburst of CN Ori) ``the remarkable feature of this diagram is the degree of continuity in the phase points, indicating that the oscillations are coherent throughout the length of the run'' (Warner \& Brickhill 1978); (from analysis of soft X-ray pulsations in SS Cyg during outburst) ``the random walk in phase model does not offer a simplified description of the data'' (Jones \& Watson 1992); ``when best defined, the DNOs show the presence of an underlying clock with abrupt changes of period, on which the phase noise seems never to cause loss of knowledge of phase'' (Warner 2004). The set of phase measurements that best displays the basic properties of DNO phase variations is that in fig.~10 of Warner \& Woudt (2002), where it can be seen that the departures from a smoothly varying phase (caused by the general increase of period during decline) are actually short sections of coherent oscillations caused by abrupt but temporary changes of period. We shall see below that there is a simple model that explains this behaviour.
   
To prepare for this we need to describe further phenomena to those already introduced in Section 2. Additional to the classic DNOs there are two kinds of quasi-periodic oscillation (QPO), one kind having periods typically fifteen times larger than the DNOs, and another type with periods $\sim$1000--3000~s (see table 1 of Warner 2004). On rare occasions double DNOs are observed, with their difference frequency equal to the QPO frequency. The LIMA model explains classic DNOs as reprocessing of the beam of high energy radiation (emerging from the accretion zone) from the accretion disc, and the additional DNO as reprocessing from a prograde traveling wave in the inner disc which is itself responsible (via obscuration and/or reprocessing of radiation from the central source) for the QPOs seen in the light curve. We therefore refer to the direct DNO as having a sidereal period, and that reprocessed from the traveling wave as having a synodic period (WW2006).  On occasion VW Hyi shows alternation between the two types of oscillation (fig.~12 of WW2006), but has other variations (frequency doubling and tripling) that complicate the situation, so we here investigate a simpler system.

\subsection{Reanalysis of the light curves of TY PsA}
The dwarf nova TY PsA was observed during a superoutburst in June 1984 and showed DNOs of sufficient amplitude for them to be seen directly in the light curve. These gave the finest opportunity to that time for following the DNO phase variations on individual nights, and lent support to the notion of decreasing coherence through outburst decline (Warner, O'Donoghue \& Wargau 1989; hereafter WOW). Here, in the light of later experience, we reinvestigate some of the DNOs present in that superoutburst and also the DNOs present in a normal outburst three months later. The relevant runs are listed in Table~3. More complete details and the light curves can be found in WOW.

\begin{table}
 \centering
  \caption{Observing runs on TY PsA}
  \begin{tabular}{llllll}
  \hline
Run    & Date (1984) & $V$  & $T$ (d)   & $t_{int}$ (s) & Duration (h) \\
  \hline
S3370  & 25 June     & 13.2 & 13               & 3                    & 3.7          \\
S3372  & 26 June     & 13.9 & 14               & 3                    & 2.7          \\
S3377  & 27 June     & 14.9 & 15               & 3                    & 0.7          \\[0.14cm]
S3409  & 22 Sept     & 13.0 & 1                & 4                    & 2.9          \\
S3412  & 23 Sept     & 13.9 & 2                & 4                    & 2.0          \\
S3412  & 24 Sept     & 14.9 & 3                & 4                    & 2.5          \\
  \hline
  \end{tabular}
\end{table}

It will be seen that the brightness of TY PsA fell at almost exactly the same rate at the ends of the two outbursts; this is a well known feature of dwarf novae -- normal and superoutbursts have nearly identical decline light curves (Warner 1995a: and see WW2002 for details of VW Hyi light curves).
  
We have analysed the light curves in the standard fashion using Fourier transforms (FTs), and also amplitude/phase ($A/\Phi$) diagrams, where we fit sinusoids, with periods derived from the FTs, by least squares to short sections of the light curve to determine the amplitudes and phases of the DNOs as functions of time.

\subsubsection{Run S3370}
In WOW the $A/\Phi$ diagram was (correctly) described as having coherent oscillations of only short duration, responsible for the noticeable broadening of the peak in the FT, compared with DNOs of longer coherence in three previous nights. As plotted in WOW the $A/\Phi$ diagram appears quite chaotic, but we noticed that with suitable adjustments of sections of the diagram, permitted by the $2\pi$ phase ambiguity of any point, a more systematic picture appears. We show this in Fig.~1, where $\Phi$ is measured relative to a fixed period of 30.17~s, and where we have partitioned the diagram into regions of clearly increasing or decreasing $\Phi$ and have indicated the DNO periods $P$ derived from FTs of each of those short sections. It is clear that $P$ alternates between two values -- approximately 30~s and 31~s, and it does so on a timescale $\sim 0.011\,\mathrm{d}$, which is close to the beat period $\sim 950\,\mathrm{s}$ between the alternating values. An FT of Fig.~1 shows a strong peak at $\sim 500\,\mathrm{s}$, with a sub-harmonic at $\sim 1000\,\mathrm{s}$.  Much of the $\Phi$ variation in Fig.~1 is therefore describable by a switching between a sidereal period $\sim 30\,\mathrm{s}$ and a synodic period $\sim 31\,\mathrm{s}$, and back again, at the beat period between the two. The similarity to the behaviour of VW Hyi, mentioned above, is evident.

\begin{figure}
 \includegraphics[width=84mm]{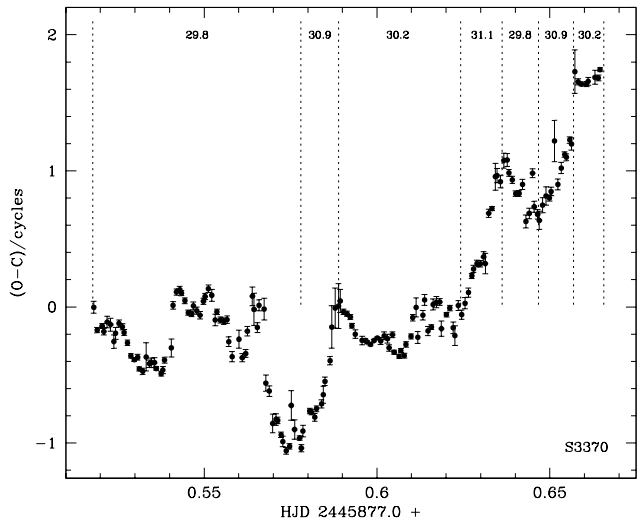}
 \caption{The phase diagram of run S3370 of TY PsA.  Numbers along the top of the diagram are the DNO periods (in seconds) measured for the sections of the light curve indicated by dotted vertical lines.  The time ranges of these sections were defined so that times when $O-C$ is systematically increasing are isolated.
}
 \label{fig:s3370omc}
\end{figure}

\subsubsection{Run S3372}
In WOW it was stated that Run S3372 does not contain any DNO signal, which, if it were the case that DNOs were really present, was attributed to a probable complete loss of coherence and hence invisibility in the FT. However, using an $A/\Phi$ search, with a DNO period initially extrapolated from the $P$ trend on previous nights, has revealed that there is one section of the light curve, lasting for almost exactly an hour, that does have a detectable signal. This is shown as an $A/\Phi$ diagram in Fig.~2, with reference period 38.0~s, where it is seen to be another example of alternation between two distinct periods, in this case $\sim 36.96\,\mathrm{s}$ and $\sim 39.93\,\mathrm{s}$, which have a beat period $\sim 500\,\mathrm{s}$, similar to the previous night, but now the switching time scale has doubled to $\sim 1000$~s.

\begin{figure}
 \includegraphics[width=84mm]{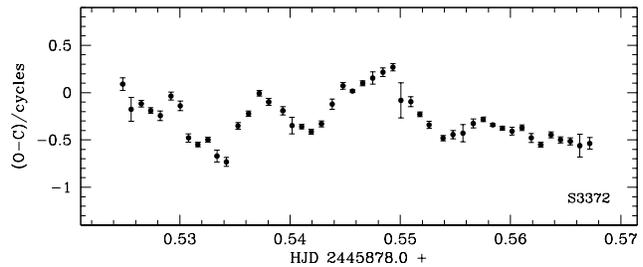}
 \caption{The $O-C$ diagram of a $\sim 1\,\mathrm{h}$ section of run S3372.  $O-C$ can be seen to alternate between increasing and decreasing.  
}
 \label{fig:s3372omc}
\end{figure}

\subsubsection{Run S3377}
WOW did not detect any DNOs in this short run, and it is true that an FT of the whole run does not show any convincing signal, but more detailed investigation shows an initial short-lived presence of quite high amplitude but uncertain period, and a persistent DNO at 47.0~s of low amplitude in the  second half of the run. At this stage TY PsA was still about 1 mag above its quiescent magnitude.

\subsubsection{Run S3409}
In WOW it was noted that this light curve possesses a DNO of large amplitude with a period of 26.5~s. Its properties are similar to those seen for run S3365 in fig.~13 of WOW and are characteristic of the early outburst decline, rather than late decline, and are not of particular interest here.

\subsubsection{Run S3412}
WOW noticed only a QPO with period $\sim 245\,\mathrm{s}$ in this run. However, we find that there is also a DNO present that changes its period so rapidly that its peak in the FT of the raw light curve is of low amplitude and greatly broadened, as seen in the FT displayed in the left hand panel of Fig.~3. The DNO is clear in the $A/\Phi$ diagram (right hand panels of Fig.~3), where the reference period is 38.59~s, but we have not found it possible to find an unambiguous interpretation of the variations. What is evident, though, are at least four large peaks of amplitude, spaced roughly 500~s apart and plotted as larger symbols in Fig.~3, and concomitant rapid switches between different periods of roughly 38.6~s and a period probably about 2 s longer. We have the impression that while the $\sim 500\,\mathrm{s}$ modulation (at twice the QPO period) is present the behaviour is similar to what is seen in Fig.~2.

\begin{figure*}
 \includegraphics[width=178mm]{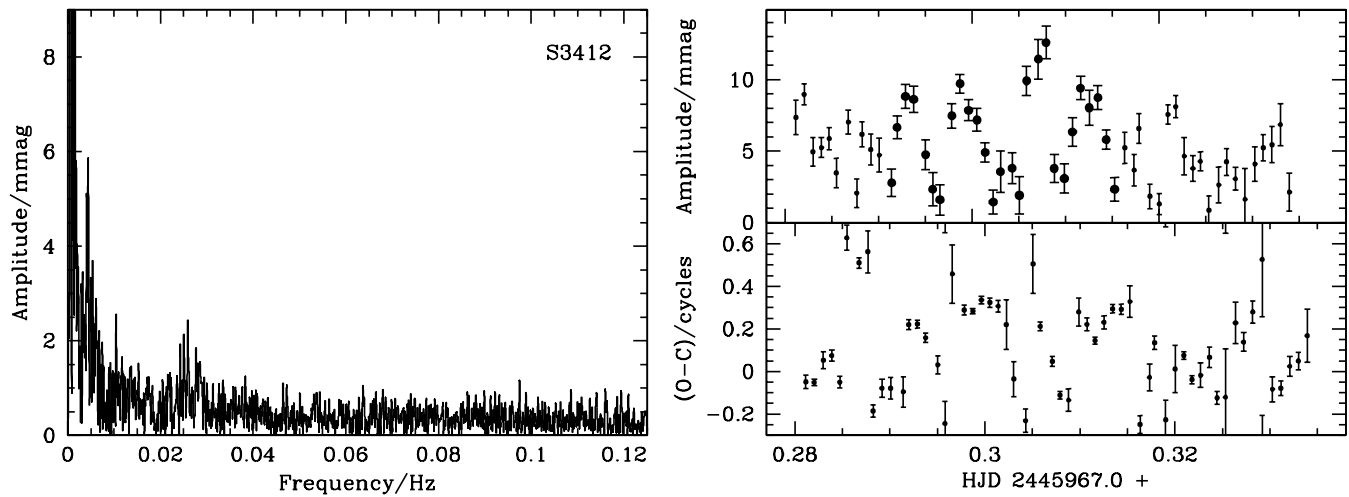}
 \caption{The FT (left hand panel) and $O-C$ and amplitude curves (right hand panels) of run S3412 of TY PsA.  $O-C$ is calculated relative to a fixed period of 38.59~s.  The DNO shows up in the FT as excess power between roughly 0.023 and 0.030~Hz.  Large symbols are used to emphasise four pulses in the amplitude curve.
}
 \label{fig:s3412}
\end{figure*}

\subsubsection{Run S3417}
Run 3417, unremarked by WOW, has a QPO with period $\sim 241\,\mathrm{s}$, which is similar to that seen on the previous night. We do not detect any DNOs in this run.

\subsubsection{Discussion}
We have found that DNOs are present later in the outburst light curve of TY PsA than had previously been realized. There is abundant evidence of the kind of switching between sidereal and synodic DNOs only previously seen in VW Hyi (WW2006). On the outburst decline a QPO periodicity at $\sim 250\,\mathrm{s}$ and its multiples is present. From a different perspective, we can say that a period $\sim 1000\,\mathrm{s}$ and its first and second harmonics are favoured, which again resembles the behaviour of VW Hyi (WW2006).
  
In the light curves of TY PsA analysed here we have not found any evidence of lpDNOs, having typically four times the periods of DNOs (Warner, Woudt \& Pretorius (2003; hereafter WWP). However, they are occasionally present in this star, as shown in fig.~3 of Warner (2004).

\section{A possible connection between the multiple DNOs in WZ Sge and longer period dwarf nova oscillations}
We aim in this Section to offer a model for the site of the lpDNOs alternative to that put forward by WWP. Our approach is via a hitherto unexplained periodic modulation in the dwarf nova WZ Sge.

\subsection{Observational overview}
WZ Sge is unusual in possessing a set of DNOs even during quiescence, which show complicated behaviour during the very rare outbursts. The key phenomena are:\\[0.4cm]
Quiescence
\begin{enumerate}
\item Two optical modulations are observed, at 27.868 and 28.952~s, discovered by Robinson, Nather \& Patterson (1978). In general either one or the other modulation is present, but occasionally both are seen simultaneously.
\item No harmonics of the 27.87-s modulation are observed, but on rare occasions a first harmonic of the 28.95-s modulation is seen (Provencal \& Nather 1997). 
\item The 27.87-s signal has substantial phase noise but is a good long-term clock (Patterson 1980; Patterson et al. 1998). The long-term stability of the 28.95-s signal has not been established.
\item Hard X-ray, soft X-ray and $UV$ modulations have been observed at the 27.87-s period (Patterson et al. 1998; Skidmore et al. 1999); however, the 28.95-s periodicity is not associated with any such high energy radiation.
\item Sidebands are observed to the 27.87-s modulation, which have been used in support of a rotating magnetic model for the white dwarf in WZ Sge (Warner, Tout \& Livio 1996; Lasota, Kuulkers \& Charles 1999; Patterson et al. 1998; WW2002).
\item Patterson (1980) suggested that the 28.95-s signal could be generated from the 27.87-s signal if the latter is reprocessed from a site moving revolving progradely in orbit at the beat period between the two, i.e. with a period of 742~s. W2002 found just such a period in light curves\footnote{The statement by Welsh et al. (2003) that the 742-s periodicity is very rarely seen is not correct: the dips that it generates in the light curves of WZ Sge are very rarely absent.} of WZ Sge and associated it with the same type of traveling wave that generates QPOs in the LIMA model.
\end{enumerate}
Outburst
\begin{enumerate}
\item DNOs were observed during the 2001 superoutburst of WZ Sge, with a dominant period of $\sim 15\,\mathrm{s}$ and short-lived modulations at $\sim 6.5\,\mathrm{s}$ thirty days after maximum (Knigge et al. 2002), and $\sim 18\,\mathrm{s}$ fifty days after supermaximum (Welsh et al. 2003). The 27.87-s periodicity had not reappeared two years later (Mukai \& Patterson 2004). It was undetectable for nearly 18 years after the 1978 outburst.
\item In contrast to the latter, the 28.95-s modulation was detected by these observers with an apparently unchanged period at various times during the outburst.
\end{enumerate}

These behaviours show certain similarities to other dwarf novae during outburst -- namely, the increase in $P$ on the decline of outburst and the frequency doubling observed at high $\dot{M}$ in SS Cyg (Mauche 1998), which may be caused by changing geometry (WW2002) or by the increasing importance of higher magnetic multipoles close to the primary. Furthermore, the independence of the 28.95-s period on luminosity of the system resembles that observed for the lpDNOs, which have been interpreted as arising from accretion directly onto the primary, i.e., associated with the spin of the primary, not of its equatorial belt (WWP).
   
There are also interesting comparisons to be made with GK Per, an IP nova remnant showing dwarf nova outbursts. It has a large amplitude 351.3-s modulation in the X-ray region during outburst (Watson, King \& Osborne 1985; Hellier, Harmer \& Beardmore 2004) which persists during quiescence (Norton, Watson \& King 1988). The 351-s period is detected in optical light curves in quiescence (Patterson 1991) but not convincingly during outburst (cf. Nogami et al. 2002). Optical quasi-periodic oscillations are seen in quiescence and outburst with a mean period near 380 s (Patterson 1981, 1991; Mazeh et al. 1985; Nogami et al. 2002). A longer period QPO is seen during outburst near 5000 s, which is the beat period between the 351-s and 380-s modulations, in X-rays (Watson et al. 1985; Hellier et al. 2004) and optical emission line profiles (Morales-Rueda, Still \& Roche 1999) but not during quiescence. The 5000-s X-ray QPOs have been explained by Hellier et al. (2004) in terms of the LIMA model, with a traveling wave at the inner edge of the disc. 
  
During outbursts of GK Per there are no observed DNOs (which would be expected to progress to lower periods than 351 s if they existed). Complete absence of DNOs is a property of IPs in general and is interpreted as implying that the magnetic field on the primary is strong enough to prevent slippage of the equatorial belt (WW2002). 
    
Finally, we note that WZ Sge has probable low amplitude 27.87-s modulation in soft X-rays and even weaker in hard X-rays (Patterson et al. 1998), the DNOs in dwarf novae and nova-like systems have large amplitude soft X-ray modulation, but at most weak signals in hard X-rays, and GK Per, as with all IPs, has predominantly hard X-ray modulation.

\subsection{Generation of the 28.95-s modulation in WZ Sge}
The observational properties listed in the previous section show that WZ Sge appears to be a bridge between the standard IP and the LIMA models. In quiescence the primary's magnetic field is barely strong enough to couple the accreting surface regions to the interior, producing considerable phase noise in the 27.87-s spin modulation, but during outburst a slipping accretion zone is possible. Two possible sites suggest themselves for the generation of the 28.95-s periodicity: (a) the main body of the primary, rotating with that period, in which case 27.87~s must be the period of a permanently slipping equatorial belt, or (b) the region just outside the corotation radius in the disc. Both sites retain their rotation periods independent of the very large mass transfer variations that occur during outburst. The former is unlikely because we would expect X-ray production from accretion onto the primary as well as its equatorial belt, but no 28.95-s X-ray modulation is observed. The latter holds more promise.
   
For WZ Sge, $28.95\,\mathrm{s}/27.87\,\mathrm{s}=1.039$. At this point we mention another rapid rotator, AE Aqr, which has $P(1) = 33.078\,\mathrm{s}$ (Patterson 1979) and such a low $\dot{M}$ that almost all accretion is propellered out of the system (e.g. Eracleous \& Horne 1996). In his discovery paper Patterson (1979) pointed out that, as well as the prominent 33.08-s modulation, there were frequent appearances of a $\sim 34.6$-s modulation and its first harmonic. The more comprehensive later study (Patterson 1994) shows that the harmonic of this feature, with $P \sim 17.4\,\mathrm{s}$, is present strongly during `flaring' episodes of AE Aqr, and is only very weakly present in quiescence. Skidmore et al. (2003) also noticed a longer period (34.23~s) during a flare. The ratio $34.8\,\mathrm{s}/33.08\,\mathrm{s}=1.052$ is comparable to what is observed in WZ Sge and again may imply that activity in the vicinity of the corotation radius is generating a modulation. In the case of AE Aqr it seems certain that propellering, which begins to operate at and beyond $r_{co}$, is a dominating mechanism, so we suspect that milder propellering is in action in WZ Sge, centrifuging gas out on field lines that connect to the disc just beyond $r_{co}$, during both quiescence and outburst.
   
Propellering has already been invoked in WZ Sge, to explain the very long intervals between superoutbursts -- a result of the inner parts of the disc being cleared of gas by the primary’s magnetosphere, preventing outbursts from starting near the inner edge of the disc, and allowing large amounts of mass to accumulate in the outer regions (Matthews et al. 2007).
   
We are now ready to propose a specific model for the generation of modulations from a region near the corotation radius. Magnetic field lines connecting from the primary to gas revolving at Keplerian velocities in an annulus just beyond $r_{co}$ are twisted back and produce a retarding torque on the primary. In the annulus a blob of threaded gas is concomitantly accelerated azimuthally \emph{and radially outwards} by the field lines, the maximum effect occurring where the field (for a tilted rotating dipole) is largest, producing a radial velocity $v_{rel} \sim 4 \gamma \beta c_s$, where $\gamma$ is the ratio of the toroidal to the vertical field strengths (close to zero at $r_{co}$), $c_s$ is the thermal sound velocity and $\beta$ is of the order of unity (Aly \& Kuijpers 1990, hereafter AK).  Immediately outside $r_{co}$ the radial velocity is subsonic but twisting of the field is already underway. Eventually the threaded gas reaches supersonic velocity and shock waves begin to form. AK show that the short-lived shocks spiral outwards and remove excess angular momentum. For a tilted dipole field a group of such shocked areas will develop each side of the disc, being carried around at a period close to the Keplerian period of the annulus, not far outside of $r_{co}$, in which the shocks occur. Heating by the shocks will thicken the disc locally; it is the cessation of this heating when the field lines snap and reconnect that was hypothesized by WW2002 to be the excitation mechanism for generating the slow prograde traveling wave at the inner edge of the disc, which produces the QPOs.
     
We suggest that the two shock regions, revolving as bright regions in the disc (which individually may be viewed differently, depending on orbital inclination, optical thickness of the disc, etc.) generate the fundamental and/or harmonic of the 28.95-s and 30.5-s modulations seen respectively in WZ Sge and AE Aqr.
      
An estimate of the total luminosity $L_{sh}$ expected from the shock regions is provided by AK. Their equation (25) can be written
\begin{equation}
L_{sh} = 6.3 \times 10^{29} \mu_{31}^2 r_9^{-5} H_8c_6\,\mathrm{erg\,s^{-1}}
\end{equation}
where $r_9=r_{co}/10^9\,\mathrm{cm}$, $H_8$ is the scale height of the disc in units of $10^8$~cm, and $c_6 = c_s/(10^6\,\mathrm{cm\,s^{-1}})$.  From modelling the outbursts of WZ Sge, Matthews et al. (2007) find they require $0.5 \le \mu_{31} \le 3$.  For $M1(1) = 1.2$ (Spruit \& Rutten 1998) we have $r_9= 1.50$, and therefore $L_{sh} \sim 1 \times 10^{29}\,\mathrm{erg\,s^{-1}}$.  When compared with the accretion luminosity of WZ Sge, $L \sim 3 \times 10^{30}\,\mathrm{erg\,s^{-1}}$ (Smak 1993) it is evident that a detectable weak modulated signal could well be generated from the shock regions. The period $P_{sh}$ of this signal will not be completely stable -- variations of conditions in the threading annulus will result in small changes in $v_{rel}$ and hence also in $P_{sh}$. 

We note that according to equation (6) $L_{sh}$ may expected to be observable only in systems with low accretion luminosity, as is the case for WZ Sge ($\dot{M} \sim 2 \times 10^{15}\,\mathrm{gs^{-1}}$; Smak 1993) and AE Aqr ($\dot{M} \sim 1 \times 10^{14}\,\mathrm{gs^{-1}}$ onto the primary; Wynn, King \& Horne 1997).  At higher $\dot{M}$, $H$ only increases $\propto \dot{M}^{3/20}$ and so $L_{sh}/L_{acc}$ could be expected to fall rapidly during the outburst of a dwarf nova (or in the high state of a nova-like variable).  But the possibility of an enhanced field in the equatorial belt, leading to an increase of $\mu$ by a factor of a few (WW2002) might, from equation (6), still give rise to a detectable $P_{sh}$ signal. In that case, it is possible that all the lpDNO signals that are observed in CVs arise not from accretion zones on the body of the primary, as suggested in WWP, but from the shock regions just outside the corotation radius. These two sites may be observationally distinguishable by photometry of lpDNOs during eclipse. The nova-like EC 21178-5417 could provide this opportunity -- it has lpDNOs with periods $\sim 95$~s and deep eclipses (WWP). For $M_1(1) = 0.8$ the source of the modulation is either at $R(1) \sim 7 \times 10^8\,\mathrm{cm}$ or near $r_{co} \sim 3 \times 10^9\,\mathrm{cm}$, a difference that might be discernable during eclipse ingress or egress, though it must be kept in mind that the whole disc could be involved in the optical signals through reprocessing of radiation from the sources.

\section{Conclusions}
A principal result of this study is further evidence that the variations in period or phase of DNOs are best described by switching between two periods -- the sidereal period, determined by the rotation period of an accreted equatorial belt, and a synodic period, probably generated by reprocessing of the radiation emitted from accretion zones on the equatorial belt. This effect is only seen where the DNO amplitudes are large enough to enable detailed analysis, but may be suspected to hold in general.

We point out that propellering of accreting gas may commonly occur towards the end of dwarf nova outbursts, depending on the existence of a sufficiently strong magnetic field on the primary. We suggest that the longer period DNOs might be located at regions just beyond the corotation radius, rather than on the surface of the primary, and that previously observed but not understood periodicities in WZ Sge and AE Aqr arise from this source.

\section*{Acknowledgements}
BW's research is partly supported by the University of Cape Town; this research was carried out while he was a Visiting Professor at Southampton University. MLP is supported by the National Research Foundation of South Africa and by the University of Southampton. We are grateful to Claire Blackman for independently analyzing one of the light curves for us.

\section*{References}
Aly J.~J., Kuijpers J., 1990, A\&A, 227, 473 \\
Cannizzo J.~K., 1993, ApJ, 419, 318\\
Cordova F.~A., Chester T.~J., Mason K.~O., Kahn S.~M., Garmire G.~P., 1984, ApJ, 278, 739\\
Eracleous M., Horne K., 1996, ApJ, 471, 427\\
Ghosh P., Lamb F.~K., 1979, ApJ, 234, 296\\
Hellier C., Harmer S., Beardmore A.~P., 2004, MNRAS, 349, 710\\
Hildebrand R.~H., Spillar E.~J., Stiening R.~F., 1981, ApJ, 248, 268\\
Horne K., Gomer R., 1980, ApJ, 237, 845\\
Jones M.~H., Watson M.~G., 1992, MNRAS, 257, 633\\
Knigge C., Hynes R.~I., Steeghs D., Long K.~S., Araujo-Betancor S., Marsh T.~R., 2002, ApJ, 580, 151\\
Long K.~S., Mauche C.~W., Raymond J.~C., Szkody P., Mattei J.~A., 1996, ApJ, 469, 841 \\
Matthews O. M., Speith R., Wynn G.~A., West R. G., 2007, MNRAS, 375, 105\\
Mauche C.~W., 1998, ASP Conf. Ser., 137, 113\\
Mauche C.~W., Robinson E.~L., 2001, ApJ, 562, 508\\
Mauche C.~W., Mattei J., Bateson F.~M., 2001, in Podsiadlowsi P., et al., eds, Evolution of Binary and Multiple Stars. Bormio, Italy\\
Mazeh T., Tal Y., Shaviv G., Bruch A., Budell R., 1985, A\&A, 149, 470\\
Morales-Rueda L., Still M.~D., Roche P., 1999, MNRAS, 306, 753\\
Mukai K., Patterson J., 2004, RevMexAA, 20, 244\\
Nogami D., Kato T., Baba H., 2002, PASJ, 54, 987\\
Norton A.~J., Watson,, M.~G., King A.~R., 1988, MNRAS, 231, 783\\
Paczynski B., 1978, in Zytkov A., ed., Nonstationary Evolution of Close Binaries. Polish Sci. Publ., Warsaw, p. 89\\
Patterson J., 1979, ApJ, 234, 978\\
Patterson J., 1980, ApJ, 241, 235\\
Patterson J., 1981, ApJS, 45, 517\\
Patterson J., 1991, PASP, 103, 1149\\
Patterson J., 1994, PASP, 106, 209\\
Patterson J., Jablonski F., Koen C., O'Donoghue D., Skillman D. R., 1995, PASP, 107, 1183\\
Patterson J., Richman H., Kemp J., Mukai K., 1998, PASP, 110, 403\\
Pretorius M.~L., Warner B., Woudt P.~A., 2006, MNRAS, 368, 361\\
Provencal J.~L., Nather R.~E., 1997, Ap. Sp. Libr., 214, 67\\
Robinson E.~L., Nather R.~E., 1979, ApJS, 39, 461\\
Robinson E.~L., Nather R.~E., Patterson J., 1978, ApJ, 219, 168\\
Schoembs R,. 1986, A\&A, 158, 233\\
Skidmore W., Welsh W.~F., Wood J.~H., Catal{\'a}n M.~S., Horne K., 1999, MNRAS, 310, 750\\
Skidmore W., O'Brien K., Horne K., Gomer R., Oke J.~B., Pearson K.~J., 2003, MNRAS, 338, 1057\\
Smak J., 1993, Acta Astron., 43, 101\\
Spruit H.~C., Taam R.~E., 1993, ApJ, 402, 593 (ST)\\
Spruit H.~C., Rutten R.~M.~G., 1998, MNRAS, 299, 768\\
Warner B., 1995a, Cataclysmic Variable Stars. Cambridge Univ. Press, Cambridge\\
Warner B., 1995, ASPC, 85, 343 \\
Warner B., 2004, PASP, 116, 115 \\
Warner B., Brickhill A.~J., 1978, MNRAS, 182, 777\\
Warner B., Robinson E.~L., 1972, Nature Phys. Sci. 239, 2\\
Warner B., Woudt P.~A., 2002, MNRAS, 335, 84 (WW2002)\\
Warner B., Woudt P.~A., 2006, MNRAS, 367, 1562 (WW2006)\\
Warner B., O'Donoghue D. Wargau W., 1989, MNRAS, 238, 73 (WOW)\\
Warner B., Woudt P.~A., Pretorius M.~L., 2003, MNRAS, 344, 1193 (WWP)\\
Watson M.~G., King A.~R., Osborne J., 1985, MNRAS, 212, 917\\
Welsh W.~F., Sion E.~M., Godon P., Gansicke B.~T., Knigge C., Long K.~S., Szkody P., 2003, ApJ, 599, 509\\
Woudt P. A., Warner B., 2002, MNRAS, 333, 411\\
Wynn G. A., King A. R., Horne K., 1997, MNRAS, 286, 436\\

\bsp

\label{lastpage}

\end{document}